\documentclass[pra,notitlepage,floatfix,amsmath,amssymb,superscriptaddress,
 reprint]{revtex4-2} 

\usepackage{adjustbox}
\usepackage{natbib} 
\usepackage{hyperref}

\usepackage{amsmath,amssymb,amsmath}
\usepackage{graphicx}
\usepackage{dcolumn}
\usepackage{bm}
\usepackage{color}
\usepackage{ulem}

\newcommand{\atomica}{\affiliation{Instituto Carlos I de F\'{\i}sica Te\'orica y Computacional and
      Departamento de F\'{\i}sica At\'omica, Molecular y Nuclear, Universidad de Granada, 18071
      Granada, Spain}}%
      
 \newcommand{\rioja}{\affiliation{\'Area de F\'{\i}sica, Universidad de La Rioja, 26006 Logro\~no, La Rioja, Spain}}%
 \newcommand{\hham}{\affiliation{Zentrum f\"ur Optische Quantentechnologien, Universit\"at
  Hamburg, Luruper Chaussee 149, 22761 Hamburg, Germany}}
  
\newcommand{\hhamcu}{\affiliation{The Hamburg Center for Ultrafast Imaging, Luruper Chaussee 149, 22761 Hamburg, Germany}}

\begin{document}

\title{Chaos and thermalization in a classical chain of dipoles}

\author{Rosario Gonz\'alez-F\'erez}\atomica
\author{Manuel I\~narrea}\rioja
\author{J. Pablo Salas}\rioja
\author{Peter Schmelcher}\hham\hhamcu


\begin{abstract} 
We explore  the connection between chaos, thermalization and ergodicity
 in a linear chain of $N$ interacting dipoles.
 Starting from the ground state, and considering chains of different numbers of dipoles, we introduce single site excitations with energy $\Delta K$. 
 The time evolution of the chaoticy of the system and the
energy localization along the chain is analyzed by computing, up to very long times,
the statistical average of the finite time Lyapunov exponent $\lambda(t)$ and of the
participation ratio $\Pi(t)$.
For small $\Delta K$, the evolution of  $\lambda(t)$ and
 $\Pi(t)$ indicates that the system becomes chaotic at roughly the same time
 as $\Pi(t)$ reaches a steady state.
For the largest values of $\Delta K$, the system becomes
 chaotic at an extremely early stage in comparison with the energy relaxation times.
 We find that this fact is due to the presence of chaotic breathers that keep the system far from equipartition and ergodicity.
Finally, we show
that the asymptotic values attained by the participation ratio $\Pi(t)$ fairly corresponds to thermal
equilibrium.
\end{abstract}



\maketitle

{\bf Introduction.}
The relationship between chaos, thermalization and ergodicity
in Hamiltonian systems with a large number of degrees of freedom is a topic of intense research with many intriguing open questions. Historically it was the pioneering study of Fermi, Pasta, Ulam and Tsingou (FPUT) in 1953 \cite{fput,fput2}, that initiated and opened up this field of research. Indeed, for a chain of nonlinear oscillators  excited out of the
equilibrium, FPUT
found that the expected energy
equipartition was not reached. Instead, they observed quasiperiodic energy
recurrences, which are more likely to occur in integrable systems.
Today, we know that these recurrences appear because the initial conditions  used by FPUT were chosen near time-periodic solutions showing a strong energy localization
in the normal mode space ({\sl q-breathers}) \cite{qb1,qb2,qb3}.
Since then, the possibility that even in weakly nonlinear Hamiltonian systems, thermalization
might not occur or be extremely slow due to the spontaneous appearance of nonergodic local fluctuations is a legitimate point of view.

In complex Hamiltonian systems, the unpredictable nature of a chaotic orbit might suggest that the corresponding dynamics is ergodic and therefore such an orbit describes a thermalized system.
The later implies that
the chaotic trajectory is able to explore all of the available phase space. 
However, the combined results of the Kolmogorov-Arnold-Moser (KAM) \cite{KAM1,KAM2,KAM3,KAM4} and the
 Nekhoroshev \cite{Nekho} theorems state
 that, in all weakly perturbed integrable systems
 it is always possible to find orbits that remain trapped 
 close to regular phase space regions up to very 
 long times. Furthermore, the remaining chaotic regions are connected due to
 Arnold diffusion, which means that, regardless of the time spent,  every chaotic orbit will eventually visit every chaotic phase space region.
Even though it is commonly accepted that the size  of the regular islands (e.g., the KAM regime) vanishes very fast, even exponentially, for increasing number of degrees of freedom \cite{A1123},
ergodicity and thermalization can only be
 fully developed in strongly perturbed Hamiltonian systems where there are (almost) no regular islands 
and phase space is then dominated by global chaos.
As a consequence, although chaos always appears as the fundamental precursor of thermalization in nonlinear lattices,  \cite{A851,A899,A1082,A1005,A1081,A908,A1000,A1079,A998,A997}, 
we also know that chaotic behavior is not 
always a sufficient condition to assert that a given orbit has also reached the thermalization
regime \cite{A908,A1079}.
In fact, in Refs.\cite{A908,A1000,A998,A997} we can find examples of nonlinear lattices 
where the time needed by the system to become chaotic is much shorter than
the ergodization time. In all those systems, the large difference between the two timescales is due to the presence of breather-like excitations.

\medskip
In this letter we use a linear chain of $N$ identical rigid interacting dipoles to elucidate the connection
between chaos, thermalization and ergodicity.
Starting with the system
in its ground state (GS), a certain amount of energy $\Delta K$ is given to one of the dipoles, and 
we explore the transport of that excess energy with increasing time evolution. To detect chaos, we compute
the maximal Lyapunov exponent $\lambda_1$ \cite{galgani1,galgani2,A879} as the limit
for $t\rightarrow \infty$ of the finite time Lyapunov exponent
\begin{equation}
\label{lyapunov}
\lambda(t)=\frac{1}{t} \log \frac{\|\bf w(t)\|}{\|\bf w(0)\|},
\end{equation}
where $\bf w(0)$ and $\bf w(t)$ are the deviation vector of a given trajectory at $t=0$ and $t>0$. 
For a regular orbit it tends to zero
as $\lambda(t) \sim t^{-1}$, while for
chaotic orbits it reaches asymptotically a nonzero value.
The inverse $\tau=1/\lambda_1$ is the Lyapunov time which
quantifies the time needed for the system to become chaotic.
To measure the degree of equipartition of the
initial excitation $\Delta K$, we use the
participation ratio $\Pi(t)$ \cite{A918,zampetaki},
\begin{equation}
\label{pr}
\Pi(t)=\frac{\Delta K^2}{\sum_{k=1}^N E_k(t)^2}-1,\quad
\end{equation}
with $E_k(t)$ the local energy stored in each dipole, that will be defined later. 
When the excitation is completely localized, carried by a single dipole, the value of $\Pi(t)$ is
zero, while if there is complete equipartition $\Pi(t)=N-1$.

\begin{figure}[h]
\centerline{\includegraphics[scale=0.3]{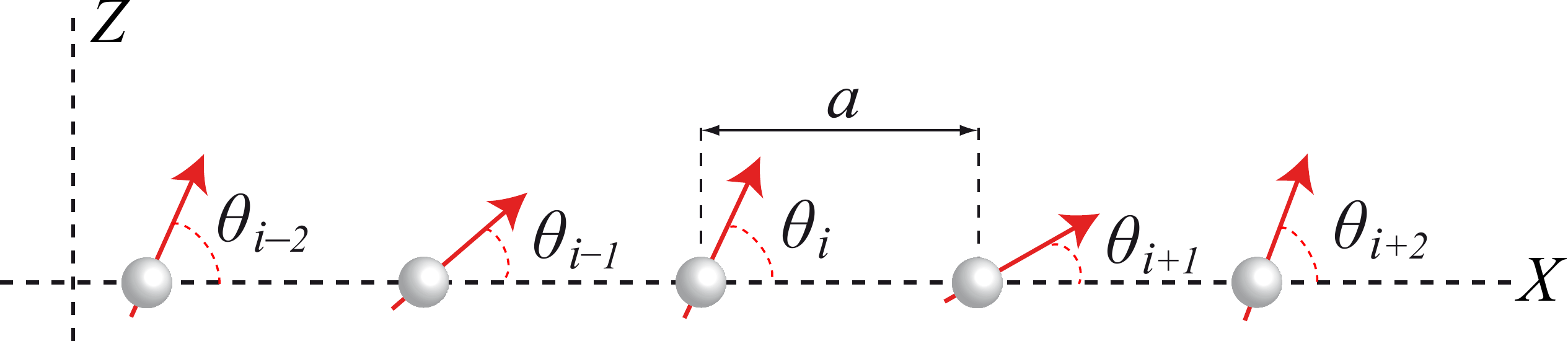}}
\caption{Schematic representation of the dipole chain.}
\label{fi:chain1}
\end{figure}

\medskip\noindent
{\bf Hamiltonian and dipole configurations.}
The dipoles are fixed in space along the $X$-axis of the Laboratory Fixed Frame $XYZ$ with a distance $a$ between two consecutive dipoles. They are restricted to rotate in the common $XZ$-plane (see Fig.\ref{fi:chain1}). Thence, the
dipole moment of each rotor is given by the vector ${\bf \mu}_i=\mu_o (\cos \theta_i, 0,\sin \theta_i)$, where $0 \le \theta_i < 2 \pi$ is the angle between the
dipole moment ${\bf \mu}_i$ and the $X$-axis, with $i=1,\dots,N$.

Assuming periodic boundary conditions (PBC) and only interactions between
nearest neighbors, the rotational dynamics of
the system, as a function of  the phases $\theta_i$, is described by the following
dimensionless Hamiltonian
\begin{equation}
\label{hamiEuler3}
{\cal H}=\sum_{i=1}^{N} \frac{p_i^2}{2} + \sum_{i=1}^{N} (\sin \theta_i \sin \theta_{i+1} - 2 \cos \theta_i \cos \theta_{i+1}),
\end{equation}
where $p_i=d\theta_i/dt$.
The energy $E={\cal H}$ in \ref{hamiEuler3} is
measured in units of $B \chi$, where  $B=\hbar^2/2 I$ is the molecular rotational constant of the dipoles, and $\chi = \mu_o^2/(4\pi\epsilon_0 a^3 B)$ is the dimensionless dipole-dipole interaction parameter in units of $B$. In this formulation, the new dimensionless time is $t'=\sqrt{\chi} \ t/t_B$ with $t_B=\hbar^2/\sqrt{2}B$. For more information about this reduction, we refer the reader to Ref.\cite{zampetaki}.

The GS of the system corresponds to the so-called head-tail
configuration $\left\lbrace \theta_i=0, \forall i\right\rbrace $ or
$\left\lbrace \theta_i=\pi, \forall i\right\rbrace $. The minimal energy of these equilibria is 
$E_m=-2 N$. Besides the GS configuration, the resulting Hamiltonian equations of motion provide us with two
families of equilibria that give rise to a complex choreography of
equilibria. One of the families is made of alternating blocks of
arbitrary number of dipoles, where all dipoles belonging to the same block are either oriented
with angles 0 or $\pi$. The other family is also made of alternating blocks of
arbitrary number of dipoles, but now with all dipoles belonging to the same block either
oriented with angles $\pi/2$ or $-\pi/2$.
Determining the nature of the equilibrium configurations involves obtaining the
eigenvalues of the stability matrix associated to \ref{hamiEuler3} and it has been achieved in \cite{zampetaki}. 
Herer, we focus on the degenerate set of equilibria given by
 only one dipole flipped with respect to the GS configuration. 
Naming these equilibria as S,  their energy is $E_s=8-2N$, and
they are saddle points. They are the equilibria
with the closest energy to the GS.
Furthermore, the energy gap between the GS and S is $\Delta_s=E_s-E_m=8$, which does not depend on the chain size. These equilibria S play a very important
role in the dynamics because, for energy values below $E_s$, the phase space
trajectories of the system remain trapped around the GS.  Conversely, for $E > E_s$, larger phase space
regions are accessible for the trajectories, which involve also different equilibria. Then, a stronger nonlinear dynamics is expected to take place. 

\medskip\noindent
{\bf Excitation dynamics.}
As we mentioned before, starting form the head-tail configuration of
minimal energy $E_m$, we excite at $t=0$ a single dipole with an excess 
energy $\Delta K$. We use chains between $N=100$ and $N=400$ dipoles. Because PBC are
assumed, without loss of generality, we excite the first dipole of the lattice.
Then, the initial conditions (i.c.) of the system are
\begin{eqnarray}
\label{ini}
\nonumber
\theta_i(0) &=& p_i(0)=0, \quad \mbox{for}  \quad i=2,...,  N,\\
\Delta K&=&\frac{p_1(0)^2}{2} + 4(1-\cos \theta_1(0)).
 \end{eqnarray}
Hereafter the energy of the system will be refered with respect to the GS energy, i.e., the total
energy of the system will be shifted by $E_m=-2 N$.
Because the energy gap $\Delta_s=8$ between the GS and the saddle point configurations S does not depend on the chain size, we provide the $\Delta K$ values in terms of that gap.
Then, because $\Delta_s$ is $N$-independent, for a given value of $\Delta K$, the larger the system's size is, the smaller the energy per dipole (energy density) $\epsilon=\Delta K/N$
is. In general, the influence of the system's size on the dynamics has been
studied keeping  $\epsilon$ constant and varying $N$ (see e.g. \cite{A899,A1069}).

For particular values of $\Delta K$, we estimate $\lambda(t)$ and $\Pi(t)$ by the simultaneous numerical integration
of the Hamiltonian equations of motion arising from \ref{hamiEuler3} and the
corresponding variational equations. More specifically, for each value of  $\Delta K$, $\lambda(t)$ and $\Pi(t)$ are statistically determined by averaging over 20
different realizations compatible with the i.c. \ref{ini}.
For the integration, we use the SABA$_2$ symplectic integrator \cite{A1095,A908}
with fixed integration time step, and with a convenient extension of the algorithm for the simultaneous integration of the variational equations.
We use an integration time step $h=0.1$ which keeps the relative energy error less than $10^{-4}$. Because the computations of $\lambda(t)$ (and so $\Pi(t)$)  require very long integration times, the code
has been parallelized.
From Hamiltonian \ref{hamiEuler3}, the local energies $E_k(t)$ appearing in \ref{pr} are defined as
\begin{eqnarray}
\label{elocal}
\nonumber
E_k(t)=\displaystyle \frac{p_k(t)^2}{2}+\frac{1}{2} \bigg[\sin \theta_k(t) \big[\sin \theta_{k+1}(t)+\sin \theta_{k-1}(t)\big] \\
- 2 \cos \theta_k(t)  \big[\cos \theta_{k+1}(t) + \cos \theta_{k-1}(t)\big] \bigg] + 2.
\end{eqnarray}
\begin{figure}[t]
\centerline{\includegraphics[scale=0.15]{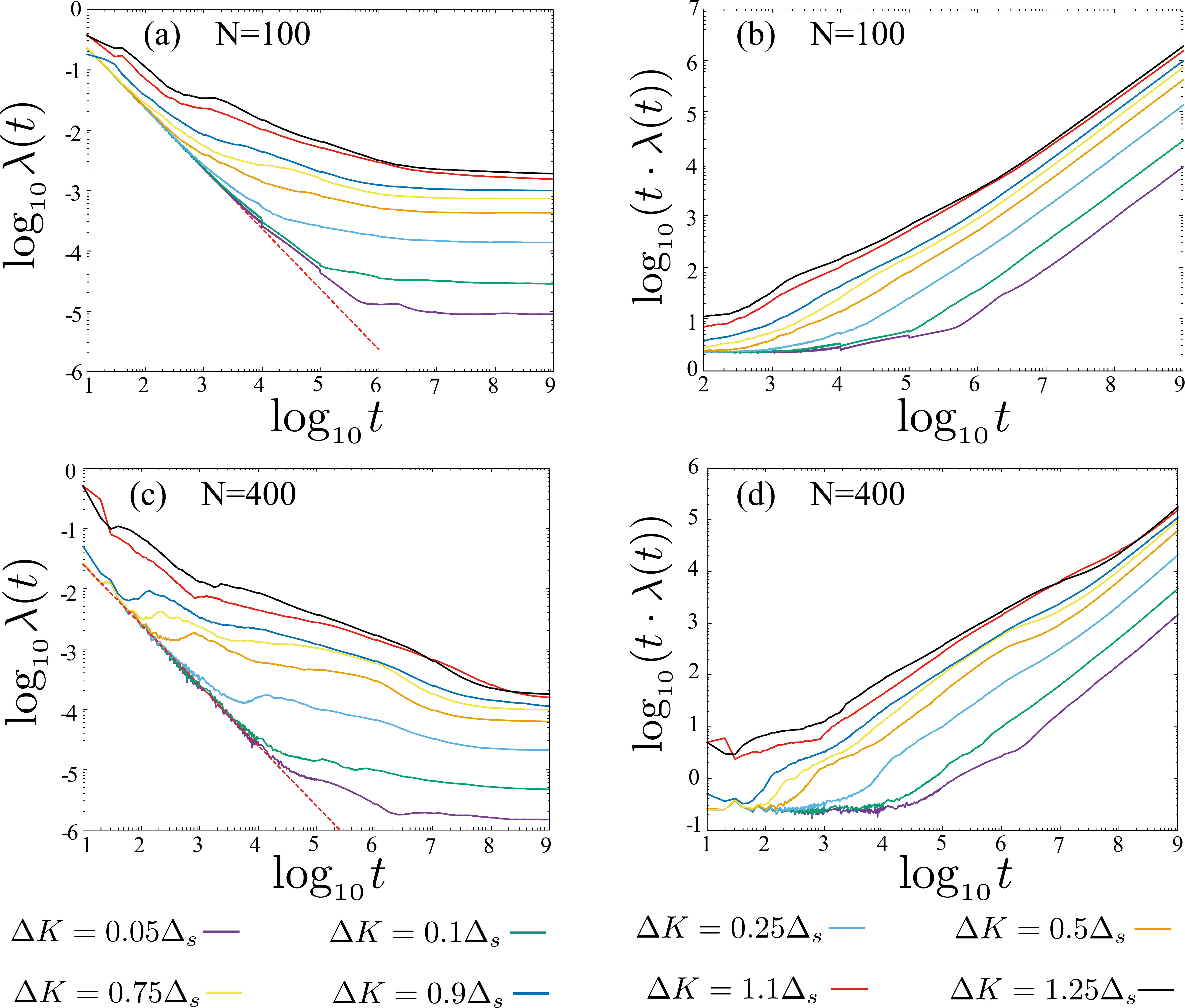}}
\caption{Left panels: Computed averaged $\lambda(t)$
for eight excess energies. The red dashed line in the left panels guides the eye for the -1 slope, which marks the expected
$\lambda(t) \sim t^{-1}$  behavior for regular
orbits. Right panels: Computed averaged $t \cdot \lambda(t)$ for the same eight excitations $\Delta K$.}
\label{fi:mlce1}
\end{figure}
\noindent
We have performed calculations
for six excitations with
 excess energy $\Delta K$ below the energy gap  $\Delta_s$, and two excitations with $\Delta K$ above  $\Delta_s$, namely
for $\Delta K=0.05\Delta_s$, $0.1\Delta_s$, $0.25\Delta_s$, $0.5\Delta_s$, $0.75\Delta_s$, $0.9\Delta_s$, $1.1\Delta_s$, and $1.25\Delta_s$.
Note that, when these excitations are below
the energy gap $\Delta_s$, the dipoles cannot perform complete rotations.

The averaged finite time Lyapunov exponent $\lambda(t)$ for two chains with $N=100$ and $N=400$ and for the above excitations are shown
in the left panel of Fig.\ref{fi:mlce1} on a double logarithmic scale. In all cases, we observe that the time evolution of $\lambda(t)$ qualitatively
shows always the same behavior. Indeed, after the system is excited, there is a transient during which
$\lambda(t)$ decreases in time. After that transient, there
is a crossover to a plateau, and the corresponding maximal Lyapunov exponent $\lambda_1$ is achieved.
However, time scales in the behavior of $\lambda(t)$ are very different depending on the value of $\Delta K$, so that
the larger the excess energy $\Delta K$ is, the shorter the transient is, and the larger the value $\lambda_1$ is.
This hierarchy in the decay patterns of $\lambda(t)$ (and so in the values of $\lambda_1$) observed in 
the left panels of Fig.\ref{fi:mlce1}
is the manifestation of an increasingly chaotic dynamics for increasing values of $\Delta K$.

After the system is excited with small and medium excess energies ( i.e., for $\Delta K=0.05\Delta_s$, $0.1\Delta_s$, $0.25\Delta_s$, $0.5\Delta_s$), the corresponding
 decay pattern of the finite time Lyapunov exponent $\lambda(t)$ (see left panels in Fig.\ref{fi:mlce1})
 closely follows the well-known power law  $\lambda(t) \sim t^{-1}$ of regular orbits.
Then, at a given time, $\lambda(t)$ 
separates from the regular behavior, and it tends to converge
to a nonzero value which is the corresponding maximal Lyapunov exponent $\lambda_1$.
This behavior, that reveals the chaotic nature of the
excitations even for very small values of $\Delta K$, 
 has been already found in different kinds of lattices such as
the FPUT problem \cite{A899,A1077,A1069}, disordered lattices
\cite{A908,A1079} or in the Bose-Hubbard model \cite{A1082}. In all these systems, including
this dipole chain, a possible explanation of the decay pattern of $\lambda(t)$ could be the existence of regions close to regular
regions in phase space where, after the initial excitation, the trajectory remains
trapped possibily for a long but finite time (given by $\tau=1/\lambda_1$)
before entering the chaotic component of the
phase space. As it was pointed out in \cite{A899,A1077}, this behavior is theoretically
sustained in the KAM and Nekhoroshev theorems \cite{KAM1,KAM2,KAM3,KAM4,Nekho}.
For $\Delta K=0.75 \Delta_s$, the regular decay of $\lambda(t)$ reduces to a very short time after
the excitation, so that for larger values of the excess energy, no trace of regular behavior
can be found in the time evolution of $\lambda(t)$.

Following \cite{A1069}, the behavior of $\lambda(t)$ in many systems can be quantitatively described by the expression
\begin{equation}
\label{lambdaeq}
\lambda(t) \approx \frac{1}{t} \log [1+ h t + c (e^{t/\tau}-1)],
\end{equation}
where $h \ll 1$ and $c$ are positive constants.
According to  Eq.\ref{lambdaeq}, in the short time regime (i.e., for $t<\tau$), $\lambda(t)$
behaves roughly linearly so that $\lambda(t) \approx \log [1+ h t ]/t$.
On the other side, in the asymptotic limit (i.e. for $t \gg \tau$), Eq.\ref{lambdaeq} converges to $\lambda(t) \rightarrow \lambda_1$.
Keeping in mind Eq.\ref{lambdaeq}, in the particular case of our dipole chain, we determine the quantity $t \cdot \lambda(t)$, which is shown in the right panel of Fig.\ref{fi:mlce1} on a double logarithmic scale.
For short times  and for small and medium excess energies (i.e., up to $\Delta K=0.5\Delta_s$), Fig.\ref{fi:mlce1}(c)-(d) show a plateau in the course of which the system behaves
roughly regularly according to $t \cdot \lambda(t) \approx \log [1+ h t ]$. For the large excitations
$\Delta K \ge 0.75\Delta_s$, there is no plateau (or it is very short)
in the curves of the right panel of Fig.\ref{fi:mlce1}, which indicates that, for these excitations, the corresponding trajectories behave chaotically from the very beginning. 

For longer times and all excess energies,  the quantity $t \cdot \lambda(t)$ increases with time. In the asymptotic limit,  this quantity 
tends to converge to a linear behavior given by
$t \cdot \lambda(t) \approx \log [c]+ \lambda_1 \ t$ (see Eq.\ref{lambdaeq}), which is observed in the right panel of Fig.\ref{fi:mlce1}
for large values of time.
Therefore, we obtain an accurate estimate of $\lambda_1$ (and so of the Lyapunov time $\tau=1/\lambda_1$) from the slope of the linear fitting of the quantities $t \cdot \lambda(t)$ for $t \ge 10^8$.

%
\begin{figure}[t]
\centerline{
\includegraphics[scale=0.3]{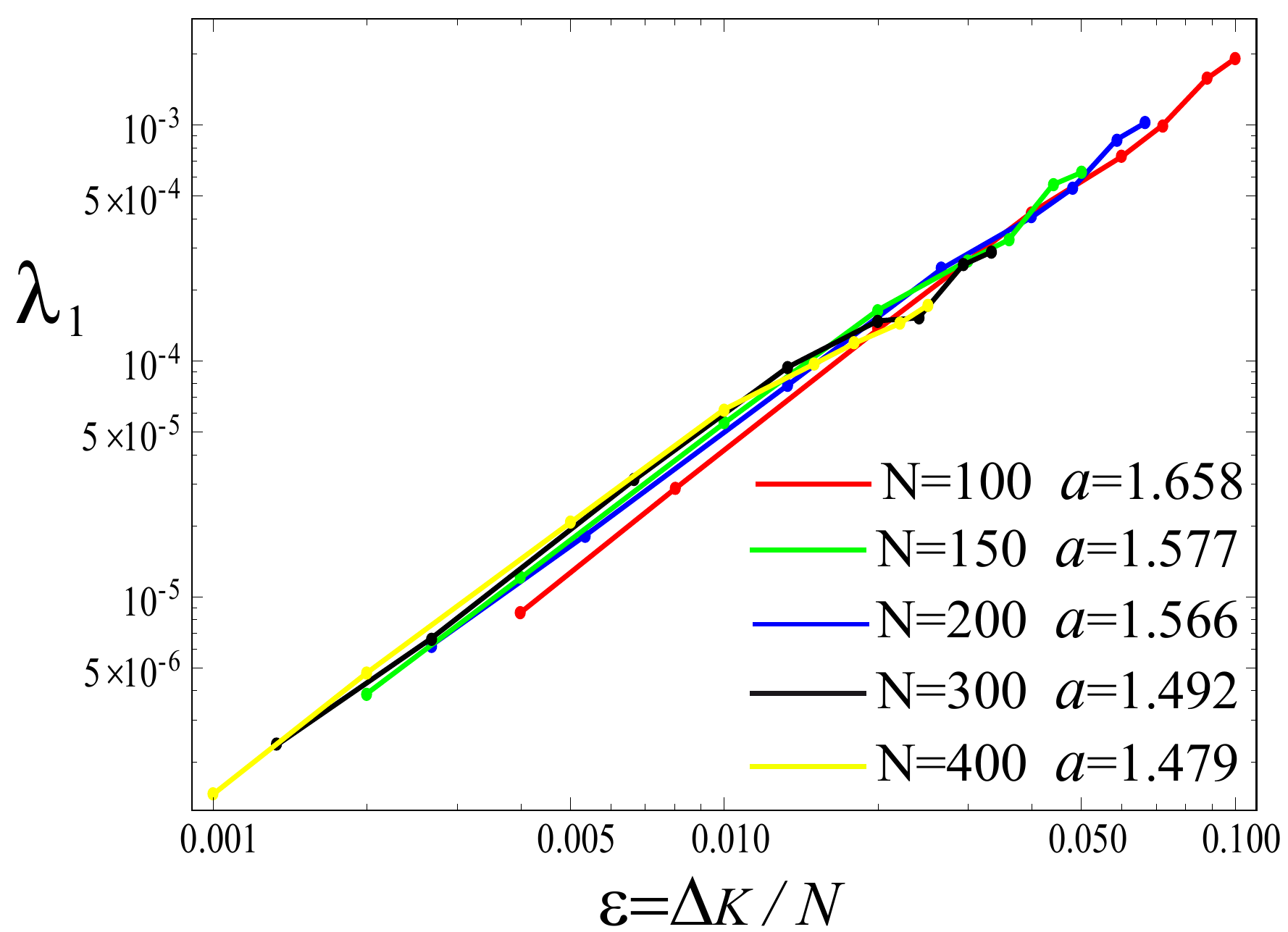}}
\caption{Maximal Lyapunov exponent  $\lambda_1$ as a function of the energy density $\epsilon=\Delta K/N$ for
dipole chains with $N=$100, 150, 200, 300 and 400. Note that a double  logarithmic
scale is used.}
\label{fi:powerlaw}
\end{figure}
For five chains of $N=100, 150, 200, 300$ and 400 dipoles, and for the eight considered excess energies $\Delta K$, we estimate
the corresponding maximal time
Lyapunov exponents $\lambda_1$ and trapping times $\tau=1/\lambda_1$ by the linear fitting
of the quantities $t \cdot \lambda(t)$  for $t \ge 10^{8}$; finding that those trapping times
are always below $\tau  \lesssim 5\times 10^5$. Furthermore, the linear behavior of 
$\lambda_1$ as a function of the energy
density $\epsilon=\Delta K/N$  shown on a double log scale in Fig.\ref{fi:powerlaw} suggests a power law
\begin{equation}
\lambda_1 \sim \epsilon^a.
\end{equation}
The least-squares fit, see Fig.\ref{fi:powerlaw}, revels a weak dependence of $a$ on $N$.
For large $N$, $a$ is expected to converge to an asymptotic value \cite{A1069}, which is not yet obtained for the $N=400$ chain analyzed here. It is important to notice that the fast decrease of
$\tau$ for increasing $\Delta K$, indicates that for low excitations, the system has
difficulties to find the {\sl gateway} to escape from the sticky quasiregular phase space regions to the non-regular counterpart.
%
\begin{figure}[t]

\centerline{\includegraphics[scale=0.15]{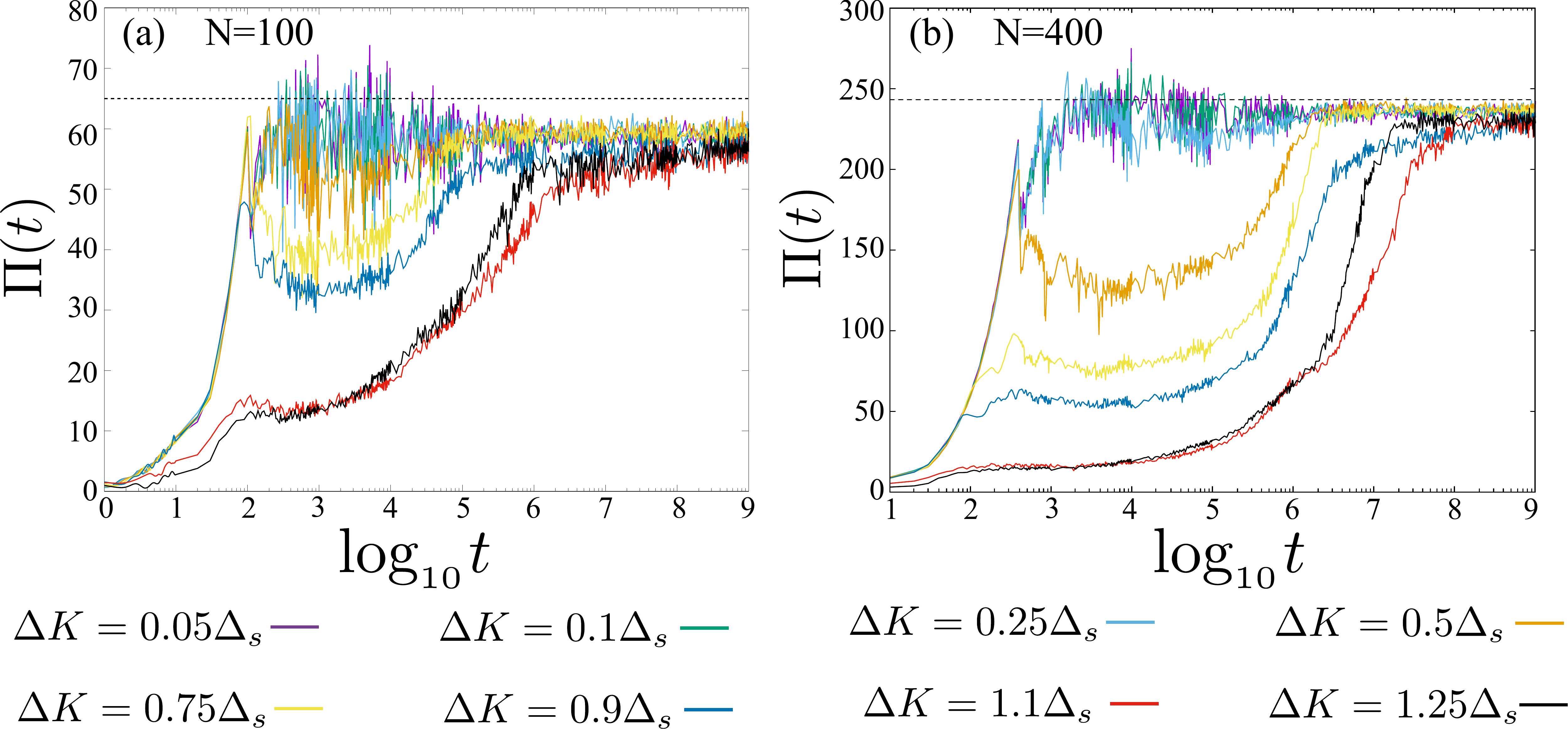}}
\caption{Time averaged participation ratio $\Pi(t)$
for eight  excess energies. The horizontal dashed lines mark the equilibrium values of $\Pi(t)$ assuming that it follows
a Boltzmann distribution in the harmonic approximation for the local energies. Chains with (a) $N=100$ and (b) $N=400$ dipoles are considered.}
\label{fi:pr1}
\end{figure}

Regarding the energy equipartition attained by the system, we ilustrate in Fig.\ref{fi:pr1} the time evolution of the averaged participation ratio $\Pi(t)$  for the same eight excess energies $\Delta K$ and for $N=100$ and 400. We observe in that figure that, for small excess energies ($\Delta K  \lesssim 0.25\Delta_s$),  there is a short
transient ($t \lesssim 100$)  during which a fast spreading of the excitation takes place.
After that transient, we always find that  $\Pi(t)$ fluctuates around a constant value. For $N=100, 150, 200, 300$ and 400, these asymptotic values are $\Pi \approx 59, 90, 118, 175$ and 238, respectively. As we can see in Fig.\ref{fi:pr1} for $N=100$ and 400, these asymptotic values are rapidly reached  for $t \gtrsim 4 \times 10^2$, and the amplitudes of the fluctuations around those constant values decrease with increasing time.
For larger excess energies, the fast initial transient in the participation ratio described for
small $\Delta K$ values is gradually replaced by a slower increase, such that $\Pi(t)$ eventually reaches asymptotic values slightly below those previously attained. Note that, for the largest excitations,
it takes longer times $t \gtrsim 10^8$  for $\Pi(t)$ to reach these constant values.

\begin{figure}[t]
\centerline{
\includegraphics[scale=0.15]{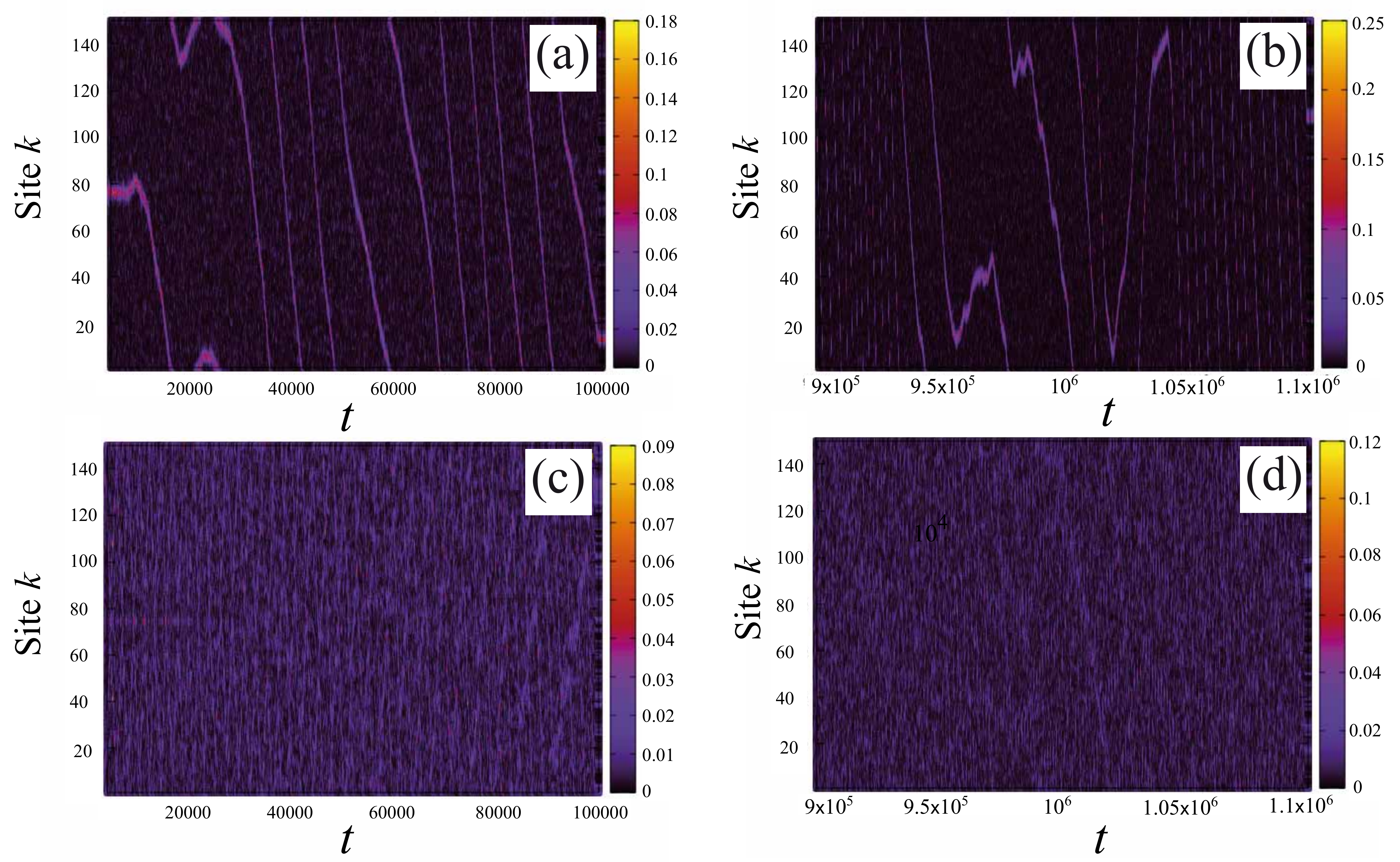}}
\caption{Color maps showing the time evolution of the local energies $E_k(t)$ (see Eq.\ref{elocal}) for a chain with $N=150$ and
$\Delta K=0.9\Delta_s$. Maps (a)-(b) correspond to an excitation with the same i.c.
but depicted for two different time intervals. Maps (c)-(d) belong to different i.c. compared to (a)-(b), but for the same time intervals.
The i.c. for trajectory showed in (a)-(b) and in (c)-(d) are $\theta_1=0$ and
$\theta_1=\pi/3$, respectively, and with the corresponding momenta
$p_1$ according to Eq.\ref{ini}.}
\label{fi:mapas}
\end{figure}

\medskip\noindent
The very long relaxation times for the largest values of $\Delta K$ indicate that, despite its chaotic dynamics
and before $\Pi(t)$ reaches the asymptotic values, the
chain exhibits a long-lasting nonergodic phase. Recent studies of 
the Gross-Pitaevskii and Klein-Gordon lattices \cite{A1000,A998} show that this nonergodic behavior is associated with
the presence of robust breather excitations that prevent the system from reaching equipartatitioning. 
For a chain of $N=150$ dipoles, the color maps of Fig.\ref{fi:mapas} show the time evolution of the local energies $E_k(t)$ (see Eq.\ref{elocal}) of two excitations with  $\Delta K=0.9 \Delta_s$, but with different i. c. $\theta_1=0$ and $\theta_1=\pi/3$ and with the momenta
$p_1$ according to Eq.\ref{ini}  (top  and bottom panels, respectively). In both cases, the time evolution of $E_k(t)$ is
shown in an early time interval (left panels) and in a much later time interval (right panels) where the dynamics has progressed substanttially.
For the excitation depicted in Fig.\ref{fi:mapas}(a)-(b), we observe that most of the energy of the system
 is strongly localized in a few energy carriers (dipoles) that
follow complex trajectories. In other words, in this case the energy transfer in the lattice is to a large extend determined by the presence of chaotic breathers \cite{A851} that keep the system far from equipartition, and exhibiting a persistent nonergodic dynamics.
However, Fig.\ref{fi:mapas}(c)-(d)
indicate that the energy transfer mechanism of the second excitation is completely different. No breather formation is observed, and even on short time scales the energy is rather distributed among all the dipoles.
The behavior shown in Fig.\ref{fi:mapas} indicates that, besides the amount of the excess energy $\Delta K$, the energy transfer mechanism is highly dependent on the way how $\Delta K$ is supplied to the
system, i.e. it dependens on  the initial conditions of the excited dipole. As a consequence, a statistical approach is necessary to obtain a general global picture of the energy transfer in the dipole chain.

Thus, we find that breathers are local hot spots that destroy the global
ergodic dynamics and therefore prevent the thermalization of the system.
It is worth noticing that this nonergodic dynamics coexists together with the global chaotic behavior that follows from the nonzero values of the maximal Lyapunov exponent $\lambda_1$.
This fact ultimately implies the lack of sensitivity of $\lambda_1$ to detect the presence of breathers, and thence to predict thermalization \cite{A1079}.
In other words, although the statistical
character of $\lambda_1$ indicates that the system exhibits a global chaotic behavior, we can not use it to assure ergodic dynamics.
A similar behavior, named
as weakly nonergodic dynamics,
was found by Mithum {\sl et al.} \cite{A998} in a Gross-Pitaevkii lattice.

\noindent
For the considered chains
the  asymptotic values of the averaged participation ratio
indicate a degree of thermalization far below the complete energy equipartition regime, for which the participation ratio
$\Pi(t)$ takes the corresponding maximum values 99, 149, 199, 299 and 399.
At this point, we pose the question of whether the
asymptotic values observed in Fig.\ref{fi:pr1} indicate that the chains are in a fairly thermalized regime.
Indeed, a numerical estimate of the equilibrium value of $\Pi(t)$ can be obtained in the following way.
Taking into account the participation ratio \ref{pr}, the estimate of its equilibrium value can be determined using the mean values of the local energy $\langle E_k \rangle$ and the squared local energy $\langle E_k^2 \rangle$ at equilibrium. Assuming that the system has a large number of dipoles and that its dynamics is ergodic, the
distribution of the local energies $E_k(t)$ (see Eq.\ref{elocal}) 
of the dipoles is governed by a Boltzmann distribution \cite{A1084,A1090}. Then, the partition function $Z$ is given by
\begin{equation}
\label{parti1}
Z =\int_{\Gamma} \exp\{-E_k(\Gamma)/T\} d\Gamma,
\end{equation}
where $\Gamma$ are the four phase variables appearing in \ref{elocal} and $T$ is the
temperature of the system at equilibrium. Thus, the mean values of the local energy $\langle E_k \rangle$ and of the squared local energy $\langle E_k^2 \rangle$ at equilibrium can be computed as  
\begin{equation}
\label{meanElocal}
\begin{array}{c}
\displaystyle{\langle E_k^i \rangle = \frac{1}{Z} \int_{\Gamma}
E_k^i(\Gamma) \exp\{-E_k(\Gamma)/T\} \ d\Gamma, }  \quad
i=1,2.
\end{array}
\end{equation}

In order to obtain the expressions of $\langle E_k \rangle$ and $\langle E_k^2 \rangle$ as functions of the temperature $T$, we calculate numerically the integrals \ref{meanElocal} for different values of $T$. These values are for a range of temperatures that correspond to the excess energy $\Delta K$ added to the system. In this way, a suitable upper limit of the temperature is estimated using the equipartition theorem and considering that the system takes the excess energy $\Delta K$ increasing only its kinetic energy. As a result, in a perfect energy equipartition regime, the mean value of the kinetic energy of each dipole is $\Delta K/N=T/2$, which provides an approximate value for the temperature $T$. Hence, taking into account the different chains and values of $\Delta K$ considered in these study, the upper limit for $T$ corresponds to the case of a chain with $N=100$ and an excess energy $\Delta K=1.25 \Delta_s$, which yields a upper limit of $T=0.2$.

\begin{figure}[t]
\centerline{
\includegraphics[scale=0.2]{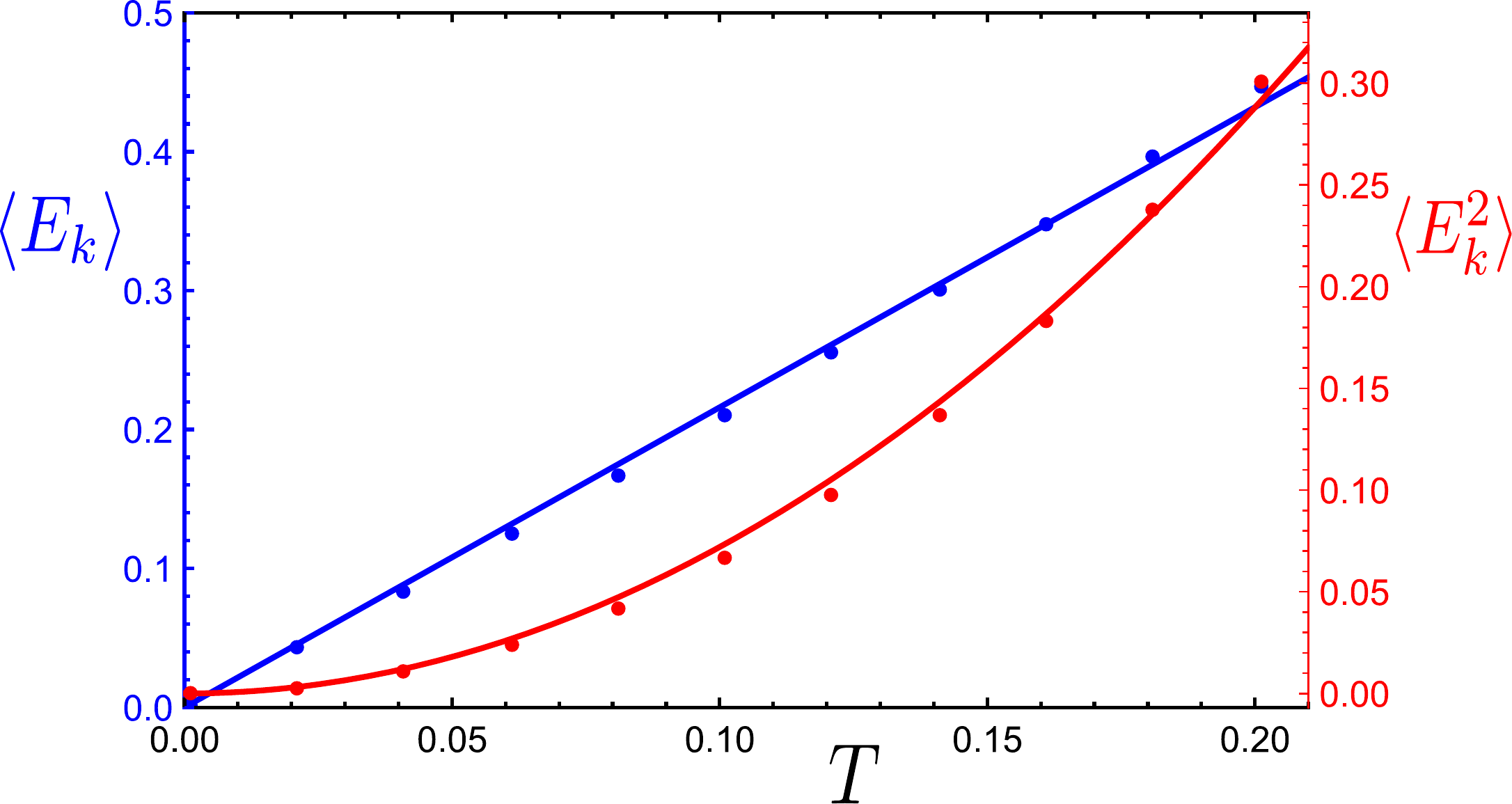}}
\caption{Evolution of the mean values of the local energy $\langle E_k \rangle$ (blue dots) and of the squared local energy $\langle E_k^2 \rangle$ (red dots) at equilibrium as functions of the temperature $T$. Blue and red lines are the least-squares linear and quadratic fitting functions.}
\label{fi:estadistica}
\end{figure}

Fig.\ref{fi:estadistica} presents the evolution of $\langle E_k \rangle$ and $\langle E_k^2 \rangle$ (see Eq.\ref{meanElocal}) as functions of $T$ in the range $0.001\le T\le 0.2$. The blue and red lines in Fig. \ref{fi:estadistica} are the least-squares linear and pure quadratic fitting functions of $\langle E_k \rangle$ and $\langle E_k^2 \rangle$ respectively. The expressions of these fitting functions are
\begin{equation}
\label{ajustes}
\langle E_k\rangle \approx 2.16 \ T, \qquad \langle E_k^2\rangle \approx 7.20 \ T^2. 
\end{equation}
Now applying these expressions in \ref{pr}, and assuming that $\Delta K=\sum_{k=1}^N E_k=N \langle E_k \rangle$ and $\sum_{k=1}^N E_k^2=N \langle E_k^2 \rangle$, the participation ratio $\Pi$ at equilibrium is given by
\begin{equation}
\label{meanPR}
\langle \Pi \rangle \approx 0.65 \ N-1.
\end{equation}
For the chains considered here, $N=$100, 150, 200, 300 and 400, we have $\langle \Pi \rangle\approx$ 64,  96, 129, 194 and  259, respectively. These values are in relatively good agreement with the asymptotic values of $\Pi$ found in Fig. \ref{fi:pr1}, being larger in all cases. As both results are rather close, it could be assumed that, once $\Pi(t)$ settles to the (still fluctuating) values observed in  Fig.\ref{fi:pr1}, the system has almost achieved thermal equilibrium. As we mentioned already, for larger excitations, the asymptotic values of $\Pi(t)$ are slightly smaller than those for small excitations, being the degree of thermalization therefore slightly smaller. However, it is clear that even in the case of very large excitations, the system is capable of reaching a degree of thermalization which is comparable to the one reached with much smaller excitations, although that requires much longer times.

Moreover, for low energy excitations it is possible to obtain analytically an approximate expression of the participation ratio $\langle \Pi \rangle$ at equilibrium. In this way, performing a series expansion of the local energies  \ref{elocal} around  the equilibrium configuation $\theta_k=p_k=0$, 
 and considering only terms up to second order in the phases $\theta_k$
(i.e., the harmonic terms), the local energies $E_k$ can be written as
\begin{equation}
\label{elocal3}
E_k(x, y, z, p_k)=\displaystyle \frac{p_k^2}{2} + 3 x_k^2 + \frac{y_k^2}{2} + \frac{z_k^2}{2},
\end{equation}
\noindent
where the new variables 
\begin{equation}
\nonumber
\Omega=(x_k, y_k, z_k, p_k)=(\theta_k/2, \theta_k/2+\theta_{k+1}, \theta_k/2+\theta_{k-1}, p_k)
\end{equation}
are defined.
Now, the partition function $Z$ \ref{parti1} is given by
\begin{equation}
\label{parti2}
Z =\int _\Omega
\exp\{-E_k(\Omega)/T\} \ d\Omega.
\end{equation}
After substituting eq. \ref{elocal3} in \ref{parti2}, we obtain
\begin{equation}
\label{parti3}
Z =\frac{4 \pi ^2}{\sqrt{6}}T^2.
\end{equation}
The mean values of $\langle E_k \rangle$ and $\langle E_k^2 \rangle$ at equilibrium read as
\begin{equation}
\label{meanElocal2}
\langle E_k^i \rangle = \frac{1}{Z} \int_\Omega
E_k^i(\Omega) \exp\{-E_k(\Omega)/T\} \ d\Omega, \quad i=1,2.
\end{equation}
These integrals can be solved analytically resulting in
\begin{equation}
\label{meanElocal1}
\langle E_k \rangle = 2 T,\qquad \langle E_k^2 \rangle =6 T^2.
\end{equation}
Following the same procedure, we get that the harmonic equilibrium value 
$\langle \Pi \rangle_H$ of \ref{pr} is given by
\begin{equation}
\label{meanPR2}
\langle \Pi \rangle_H =\frac{2 N}{3}-1.
\end{equation}
The equilibrium value $\langle \Pi \rangle_H$ is slightly larger than the equilibrium value 
$\langle \Pi \rangle \approx 0.65 \ N-1$ (see Eq.\ref{meanPR})
determined numerically using the exact expression \ref{elocal} for the local energy.
Even though only the linear terms were taken into account, the harmonic equilibrium value \ref{meanPR2} is close to the equilibrium value \ref{meanPR} to
be considered as an upper bound for
the asymptotic value of the participation function $\Pi(t)$.

\medskip
From the numerical results of the time evolution of  $\Pi(t)$
depicted in Fig.\ref{fi:pr1}, it is clear that the
larger the excitation is, the longer the time is for the system reach energy equipartition.
A rough estimate of the  thermalization times can be obtained
from  Fig.\ref{fi:pr1}. For excitations $\Delta K  \lesssim 0.75 \Delta_s$, thermal equilibrium is reached
for $t \gtrsim 10^5$, while for  $\Delta K  \gtrsim 0.75 \Delta_s$, thermalization requires times that, in many cases, 
are greater than $t=10^8$.
From these estimates of the energy equipartition time, it is clear that, except for the smallest excitation value $\Delta K  = 0.05 \Delta_s$, the
thermalization times are always
 larger than the corresponding Lyapunov times $\tau$  (see Fig.\ref{fi:powerlaw}(b)).
 Indeed, our computations show that the system becomes chaotic before an acceptable energy equipartition is achieved.
 For $\Delta K  = 0.05 \Delta_s$, the system roughly becomes
 chaotic at the same time as equipartition is achieved.
 
 {\bf Conclusions.}
 We have explored the connection between chaos, thermalization and ergodicity
 in a linear chain of hundred interacting dipoles. Starting from the GS, the chains have been excited by supplying different excess energies
 $\Delta K$ to one of the dipoles. Our tools were the finite time Lyapunov exponent $\lambda(t)$ \ref{lyapunov} and the participation ratio $\Pi(t)$ \ref{pr}, which provide information about the chaoticity of the system and the localization of the energy.


It turns out that the averaged $\lambda(t)$ shows always the same behavior: Once the system is excited, there is a transient during which
$\lambda(t)$ decreases in time. After the transient, there
is a crossover to a plateau, and the corresponding maximal Lyapunov exponent $\lambda_1$ is reached asymptotically. 
However, the value of $\Delta K$ dictates the strongly varying times scales of the behavior of
$\lambda(t)$: A larger excess energy $\Delta K$ implies a shorter transient and a larger value of $\lambda_1$.
This hierarchy indicates an increasingly chaotic dynamics for increasing values of $\Delta K$.

When the system is excited with small and medium $\Delta K$  values the
 decay pattern of  the averaged $\lambda(t)$
 closely follows the expected power law  $\lambda(t) \sim t^{-1}$ of regular orbits.
Then, at a given time, $\lambda(t)$ 
diviates from this regular behavior, and it tends to converge
to the corresponding $\lambda_1$ value.
For the largest excitation energies considered here,  there is no trace of regular behavior in the decay of $\lambda(t)$ before the corresponding asymptotic value of $\lambda_1$ is reached. 

For small excess energies,  the averaged $\Pi(t)$ shows a short
transient with a fast spreading of the excitation.
After that transient, $\Pi(t)$ fluctuates around a constant value
which depends on $N$. For larger values of $\Delta K$, the fast initial transient observed for
small $\Delta K$ values is replaced by a slow increase. Thence, for long times, $\Pi(t)$ eventually reaches an asymptotic value.

For the largest values of $\Delta K$, we found that the extremely long relaxation times showed by $\Pi(t)$ in comparison with the values of the Lyapunov times $\tau$ are due to the presence of chaotic breathers that keep the system far from equipartition.
Furthermore, we observed that, besides the value of $\Delta K$, the energy transfer mechanism is highly dependent on the i.c. of the excited dipole. As a consequence, a statistical approach as the one carried out in this paper is necessary to obtain a correct description of the energy transfer mechanicsm in the dipole chain.

The asymptotic values of the averaged $\Pi(t)$ numerically calculated 
indicate a degree of thermalization well below the energy equipartition. Assuming the ergodicity of the system at thermal equilibrium, we have determined the thermal equilibrium values of $\Pi(t)$ by means of the Boltzmann statistics.
We find that the thermal equilibrium values of $\Pi$ are in good agreement with the asymptotic values attained by $\Pi(t)$. Since both values are rather close, we can assert that the 
asymptotic values of $\Pi(t)$ indicate that the system has almost achieved thermal equilibrium, which on the other side, is far from a perfect energy equipartition regime.


A natural continuation of the present work is its extension to more complex  dipole systems, such as dimerized dipole chains and one-dimensional arrays of dipoles (e.g.,  diamond and sawtooth arrays \cite{A1021}). One exciting direction is the possibility of identifying or even desingning flat bands (see e.g. Ref.\cite{A1022} and references therein) in such one-dimensional arrays of dipoles and to study their impact on the energy transfer mechanism of the system. 

\bigskip
{\bf Acknowledgments.} M.I. and J.P.S. acknowledge financial support by the Spanish Project No. MTM 2017-88137-C2-2-P (MINECO). R.G.F. gratefully acknowledges financial support by the Spanish Project No. FIS2017-89349-P (MINECO), PY20$\_$00082 (Junta de Andalucía), and by the Andalusian research group FQM-207. This study has been partially financed by the Consejería de Conocimiento, Investigación y Universidad, Junta de Andalucía and European Regional Development Fund (ERDF), Ref. SOMM17/6105/UGR.
These work used the Beronia cluster (Universidad de La Rioja), which is supported by FEDER-MINECO grant UNLR-094E-2C-225.

\end{document}